\documentclass[twocolumn,showpacs,preprintnumbers,amsmath,amssymb]{revtex4}

\usepackage{graphicx}
\usepackage{dcolumn}
\usepackage{bm}

\begin{document}

\preprint{}

\title{Doppler- and recoil-free laser excitation of Rydberg states via three-photon transitions}
\author{I.~I.~Ryabtsev}
  \email{ryabtsev@isp.nsc.ru}
\author{I.~I.~Beterov}
\author{D.~B.~Tretyakov}
\author{V.~M.~Entin}
\author{E.~A.~Yakshina}
\affiliation{A.~V.~Rzhanov Institute of Semiconductor Physics SB RAS\\ Prospekt Lavrentyeva 13, 630090 Novosibirsk, Russia }

\date{November 18, 2011}

\begin{abstract}
Three-photon laser excitation of Rydberg states by three different laser beams can be arranged in a starlike geometry that simultaneously eliminates the recoil effect and Doppler broadening. Our analytical and numerical calculations for a particular laser excitation scheme $5S_{1/2} \to 5P_{3/2} \to 6S_{1/2} \to nP$ in Rb atoms have shown that, compared to the one- and two-photon laser excitation, this approach provides much narrower linewidth and longer coherence time for both cold atom samples and hot vapors, if the intermediate one-photon resonances of the three-photon transition are detuned by more than respective single-photon Doppler widths. This method can be used to improve fidelity of Rydberg quantum gates and precision of spectroscopic measurements in Rydberg atoms.
\end{abstract}

 \maketitle

\section{Introduction}

When an atom absorbs a photon, it obtains a momentum $\mathbf{P}=\hbar \mathbf{k}$, where $\mathbf{k}$ is a wave vector of the photon. This results in the change of the atom velocity $\mathbf{V}$ (recoil effect) and also of the Doppler shift $\omega _{D} =\mathbf{k V}$, which changes the resonance frequency for the next interaction of this atom with photons [1]. The recoil effect is negligible for hot atoms, but it may affect atom-light interaction in ultracold atom samples with temperatures $T\le 1\; \mu {\rm K}$ (e.g., in Bose-Einstein condensates). In particular, the recoil Doppler shift of the resonance frequency and Doppler broadening generally lead to decoherence at laser excitation, while the recoil effect itself causes the heating of an atom sample [2]. 

Decoherence and heating at laser excitation can be crucial for the most promising applications of ultracold atoms: quantum-information processing [3], precision spectroscopy [4], and metrology [5]. In quantum information processing with neutral atoms, two-qubit quantum gates are implemented using laser excitation of strongly interacting Rydberg states for a short time and then returning them back to the ground state [3]. High-fidelity quantum computation requires the gate error to be $\le 10^{-4} $; therefore, the recoil and Doppler effects should be as small as possible, since they contribute to the overall gate fidelity. In precision spectroscopy of ultracold atoms, these effects are among the main limiting factors in measuring the transitions frequencies at the accuracy level of $<10^{-12} $ [4,5].

In this paper we analyze in detail a Doppler- and recoil-free laser excitation of Rydberg states via three-photon transitions using three different laser beams arranged in a starlike planar geometry. This method was first proposed in Ref.~[6] and then implemented in Ref.~[7] for Doppler-free spectroscopy of the $3S_{1/2} \to 3P_{1/2} $ transition in Na atoms using three-photon induced absorption and emission with two far-detuned laser radiations. The 60-MHz-wide hyperfine resonances have been observed in a hot Na vapor cell. A possibility of the Doppler-free three-photon excitation of Rydberg states was mentioned in Ref.~[8] but it was not analyzed in detail.

\section{One- and two-photon Rydberg excitation}

We proceed with an example of Rb atoms, which are mainly used in the experiments on quantum information processing. Figure 1(a) depicts possible Rydberg excitation schemes in Rb atoms. 

One-photon excitation on the $5S_{1/2} \to nP$ transitions at 297 nm gives the recoil velocity of 1.5 cm/s and recoil Doppler shift of 51 kHz, and it is obviously unsuitable to Doppler- and recoil-free spectroscopy in ultracold atoms since there is no way to suppress the Doppler and recoil effects, except the M\"ossbauer-like effect in optical lattices [5]. In a hot Rb vapor cell the resonances of the 20$-$40 MHz width were observed on these transitions using electron-shelving-like technique [9].

Most often used is the two-photon Rydberg excitation $5S_{1/2} \to 5P_{3/2} \to nS,\, nD$ with the 780 nm radiation at the first step and 480 nm at the second step [10-12]. The 780 nm photon gives the recoil velocity of 0.59 cm/s and recoil Doppler shift of 20 kHz, while the 480 nm photon gives the recoil velocity of 0.96 cm/s and recoil Doppler shift of 32 kHz. The recoil effect and Doppler shift can be partially canceled if the counterpropagating laser beams at 780 and 480 nm are used [3], but there is still no way to completely eliminate the Doppler and recoil effects at this two-photon transition. They would be suppressed if the two-photon transition was excited by a single 594 nm radiation formed as a standing wave, due to exact compensation of the recoil and Doppler effects in two counterpropagating light waves [13]. Unfortunately, there is no intermediate state close enough to a virtual intermediate level at the 594 nm wavelength, so that too much laser power would be required to excite this transition.

\begin{figure}
\includegraphics[scale=0.6]{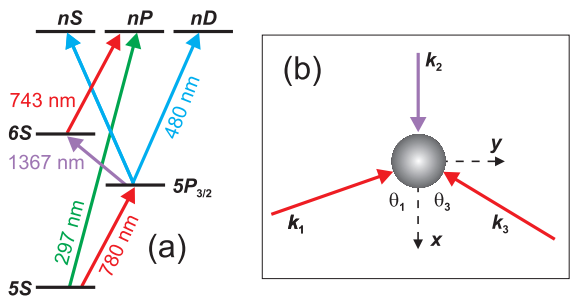}
\caption{\label{Fig1} (Color online) (a) Scheme of the one-, two-, and three-photon laser excitation of Rydberg states in Rb atoms. (b) Star-like geometry for Doppler- and recoil-free Rydberg excitation via a three-photon transition $5S_{1/2} \to 5P_{3/2} \to 6S_{1/2} \to nP$. Wave vector $\mathbf{k_{1}} $ corresponds to the 780 nm laser beam, $\mathbf{k_{2}} $ to 1367 nm, and $\mathbf{k_{3}} $ to 743 nm.}
\end{figure}

\section{Doppler- and recoil-free three-photon Rydberg excitation}

We now turn to a three-photon Rydberg excitation scheme $5S_{1/2} \to 5P_{3/2} \to 6S_{1/2} \to nP$, with 780 nm radiation (wave vector $\mathbf{k_{1}} $) at the first step, 1367 nm  (wave vector $\mathbf{k_{2}} $) at the second step, and 743 nm  (wave vector $\mathbf{k_{3} }$) at the third step [Fig.~1(a)]. We have launched this scheme recently in our experimental setup with a Rb magneto-optical trap. Similar three-photon schemes have also been implemented by the other groups in Rb vapor cells [14-16]. According to Ref.~[6], the three exciting laser beams $\mathbf{k_{1}} $, $\mathbf{k_{2}} $, and $\mathbf{k_{3}} $ can be sent to atoms from various directions, and we may expect that some specific geometry exists that zeros the recoil effect and satisfies the condition

\begin{equation} \label{Eq1} 
\mathbf{k_{1}} +\mathbf{k_{2}} +\mathbf{k_{3}} =0.   
\end{equation} 

For our three-photon transition such geometry is sketched in Fig.1(b). The three laser beams are intersecting at the point of a small atom sample. For distinctness, we suppose that all three wave vectors lie in the \textit{x-y} plane with the \textit{x} axis directed along  $\mathbf{k_{2}} $ (planar star-like geometry, which is most convenient for experiments). Wave vectors $\mathbf{k_{1}} $ and $\mathbf{k_{3}} $ have the incidence angles $\theta _{1} $ and $\theta _{3} $ with respect to the \textit{x} axis. In such a geometry Eq.~\eqref{Eq1} gives for unknown angles $\theta _{1} $ and $\theta _{3} $

\begin{equation} \label{Eq2} 
\begin{array}{l} {k_{1} \sin \theta _{1} =k_{3} \sin \theta _{3} \; ,} \\ \\{k_{1} \cos \theta _{1} +k_{3} \cos \theta _{3} =k_{2} \; .} \end{array} 
\end{equation} 

\noindent Here $k_{j} =2\pi /\lambda _{j} $ are respective wave numbers of each light wave. From Eq.~\eqref{Eq2} we find

\begin{equation} \label{Eq3} 
\begin{array}{l} {\cos \theta _{1} =(k_{1}^{2} +k_{2}^{2} -k_{3}^{2} )/(2k_{1} k_{2} )\; ,} \\ \\ {\cos \theta _{3} =(-k_{1}^{2} +k_{2}^{2} +k_{3}^{2} )/(2k_{2} k_{3} )\; .} \end{array} 
\end{equation} 

\noindent Equation~\eqref{Eq3} yields $\theta _{1} \approx 78.7{}^\circ $ and $\theta _{3} \approx 69.1{}^\circ $. This confirms that the proposed geometry of Fig.~1(b) can be implemented for recoil-free laser excitation of Rydberg states. 

We now consider the Doppler effect for a single atom having arbitrary velocity vector \textbf{V} with components $V_{x} ,\; V_{y} ,$ and $V_{z} $ and interacting with the three laser beams in the geometry of Fig.~1(b). The Doppler shifts at each intermediate one-photon transition are

\begin{equation} \label{Eq4} 
\begin{array}{l} {\Delta \omega _{1} =\mathbf{k_{1} V}=-k_{1} V_{x} \cos \theta _{1} +k_{1} V_{y} \sin \theta _{1} \; ,} \\ \\ {\Delta \omega _{2} =\mathbf{k_{2} V}=k_{2} V_{x} \; ,} \\ \\ {\Delta \omega _{3} =\mathbf{k_{3} V}=-k_{3} V_{x} \cos \theta _{3} -k_{3} V_{y} \sin \theta _{3} \; .} \end{array} 
\end{equation} 

\noindent If we substitute Eq.~\eqref{Eq2} into Eq.~\eqref{Eq4}, the overall Doppler shift of our three-photon transition turns out to be exactly zero:

\begin{equation} \label{Eq5} 
\Delta \omega =\Delta \omega _{1} +\Delta \omega _{2} +\Delta \omega _{3} \equiv 0\; . 
\end{equation} 

\noindent Independently of \textbf{V} the overall Doppler shift is automatically zeroed in the geometry of Fig.~1(b) if angles $\theta _{1} $ and $\theta _{3} $ are chosen according to Eq.~\eqref{Eq3}. This means that three-photon transitions and starlike planar geometry of the three laser beams not only provide a recoil-free excitation of Rydberg states, but can also provide a Doppler-free excitation in atom samples with arbitrary velocity distributions. We may expect that this approach will work both for ultracold atom samples in optical and magnetic traps and for hot atoms in vapor cells.

\section{Theory for three-photon Rydberg excitation}

In order to confirm the above conclusions, we need to calculate the spectra of three-photon excitation for atom samples of various temperatures. Several assumptions should be made on a theoretical model to be used. Both in quantum information processing and in high-resolution spectroscopy, the linewidth of the three-photon transition must be as small as possible. If all intermediate one-photon transitions are on resonance, the line width will be a few megahertz, since the natural widths of the $5P_{3/2} $ and $6S_{1/2} $ states are 6 and 3 MHz, respectively. For example, in the vapor-cell experiments on resonant three-photon excitation the observed linewidths were 10$-$20~MHz [14-16]. In order to reduce the line width, the intermediate states should be far detuned from the laser radiations at each one-photon transition, that is, one-photon detunings should be much more than respective one-photon Rabi frequencies. In this case populations of the intermediate states are close to zero; we may neglect spontaneous decay from these states and use the Schr\"odinger's equation instead of the density matrix in calculations. The ultimate linewidth of the three-photon transition in a frozen atom is defined either by an inverse lifetime $\tau_R$ of the final Rydberg state (2$-$3 kHz natural width for the principal quantum numbers $n\sim 50$ [17]) or by an inverse interaction time of atoms with laser radiation (Fourier-transform width).

In the further calculations we denote the $5S_{1/2} $ state to be state 0, $5P_{3/2} $ to be state 1, $6S_{1/2} $ to be state 2, and the $nP$ Rydberg state to be state 3. For each intermediate one-photon transition $j=1,\; 2,\; 3$ we specify the respective Rabi frequencies $\Omega _{j} =d_{j} E_{j} /\hbar $ (here $d_{j} $ are dipole moments of one-photon transitions and $E_{j} $ are electric-field amplitudes of the linearly polarized light fields) and detunings $\delta _{j} $, which also include the Doppler shifts. The total detuning $\delta =\delta _{1} +\delta _{2} +\delta _{3} $ from the three-photon transition $0\to 3$ is independent of the atom velocity due to cancellation of the Doppler shift according to Eq.~\eqref{Eq5} and is supposed to be small, while $\delta _{j} $ are supposed to be large in order to not populate intermediate states 1 and 2. The Schr\"odinger equation gives for the probability amplitudes $a_{j} $ of states $j=0-3$ in the rotating wave approximation

\begin{equation} \label{Eq6} 
\begin{array}{l} {\dot{a}_{0} =-i\Omega _{1} a_{1} e^{i\delta _{1} t} /2,} \\ \\ {\dot{a}_{1} =-i(\Omega _{1} a_{0} e^{-i\delta _{1} t} +\Omega _{2} a_{2} e^{i\delta _{2} t} )/2,} \\ \\ {\dot{a}_{2} =-i(\Omega _{2} a_{1} e^{-i\delta _{2} t} +\Omega _{3} a_{3} e^{i\delta _{3} t} )/2,} \\ \\ {\dot{a}_{3} =-i\Omega _{3} a_{2} e^{-i\delta _{3} t} /2.} \end{array} 
\end{equation} 

Amplitudes $a_{0} $ and $a_{3} $ are slowly varying variables at the three-photon Rabi frequency, while $a_{1} $ and $a_{2} $ are small but rapidly oscillating values due to large detunings $\delta _{1} $ and $\delta _{2} $. We therefore can make the replacements $a_{1} =\alpha _{1} e^{-i\delta _{1} t} $ and $a_{2} =\alpha _{2} e^{-i\delta _{3} t} $ in order to decompose them into slowly varying parts $\alpha _{1,\; 2} \to 0$ and rapidly oscillating exponents. Substituting these replacements into Eq.~\eqref{Eq6} and neglecting the small terms containing $\dot{\alpha }_{1} $ and $\dot{\alpha }_{2} $, we finally obtain the population of Rydberg state at three-photon laser excitation

\begin{equation} \label{Eq7} 
\begin{array}{l} {\left|a_{3} \right|^{2} \approx \displaystyle \frac{\Omega ^{2} }{\Omega ^{2} +(\delta +\Delta _{0} +\Delta _{3} )^{2} } } \\ {\times \displaystyle \frac{1}{2} \left[1-\cos \left(t\sqrt{\Omega ^{2} +(\delta +\Delta _{0} +\Delta _{3} )^{2} } \right)\right]\; ,} \end{array} 
\end{equation} 

\noindent where $\Omega =\Omega _{1} \Omega _{2} \Omega _{3} /(4\delta _{1} \delta _{3} )$ is the three-photon Rabi frequency, and $\Delta _{0} =\Omega _{1}^{2} /(4\delta _{1} )$ and $\Delta _{3} =\Omega _{3}^{2} /(4\delta _{3} )$ are the power shifts of states 0 and 3, respectively. The power shifts occur at transitions $0\to 1$ and $2\to 3$ due to the interaction with the far-detuned laser radiations. More sophisticated density matrix calculations, which take into account a slow decay of the Rydberg state at rate $\gamma =1/\tau _{R} \ll\Omega $, give a similar formula,

\begin{equation} \label{Eq8} 
\begin{array}{l} {\rho _{33} \approx \displaystyle \frac{\Omega ^{2} }{\Omega ^{2} +(\delta +\Delta _{0} +\Delta _{3} )^{2} +\gamma ^{2} /4} } \\ {\times \displaystyle \frac{1}{2} \left[1-e^{-\gamma t/2} \cos \left(t\sqrt{\Omega ^{2} +(\delta +\Delta _{0} +\Delta _{3} )^{2} } \right)\right]\; .} \end{array} 
\end{equation} 

\noindent It exhibits a slow dephasing at long excitation times (comparable with $\tau _{R} $), which is negligible on a microsecond time scale for $n\ge 50$. Several observations can be made from Eq.~\eqref{Eq8}. 

First, the three-photon excitation spectrum can be recorded by varying any of $\delta _{1} ,\; \delta _{2} ,\; {\rm or}\; \delta _{3} $, provided $\delta =\delta _{1} +\delta _{2} +\delta _{3} $ is small and satisfies the resonance condition. Varying $\delta _{2} $ would be a good choice because power shifts $\Delta _{0} $ and $\Delta _{3} $ are independent of $\delta _{2} $. Moreover, there is no requirement for $\delta _{2} $ to be as large as $\delta _{1} $ and $\delta _{3} $, since in Eq.~\eqref{Eq8} $\delta _{2} $ appears only in $\delta $. It can even be zero, while $\Omega _{2} $ can be large (tens of megahertz) in order to make $\Omega $ as large as possible and excite three-photon transition on a short (microsecond) time scale. At the same time, $\Omega _{1} $ and $\Omega _{3} $ should satisfy $\Omega _{1} \ll\delta _{1} $ and $\Omega _{3} \ll\delta _{3} $ in order for Eq.~\eqref{Eq8} to be valid and no intermediate states are populated via one-photon transitions.

Second, the Doppler shift is canceled in $\delta $ but it is present in $\delta _{1} $ and $\delta _{3} $, and correspondingly in $\Delta _{0} $ and $\Delta _{3} $. The power shifts depend on the atom velocity \textbf{V} as $\Delta _{0} =\Omega _{1}^{2} /[4(\delta _{1}^{*} -\mathbf{k_{1} V})]$ and $\Delta _{3} =\Omega _{3}^{2} /[4(\delta _{3}^{*} -\mathbf{k_{3} V})]$, where $\delta _{1}^{*} $ and $\delta _{3}^{*} $ are laser detunings for a frozen atom. In an atom sample of temperature \textit{T} the one-dimensional Maxwell velocity distributions are

\begin{equation} \label{Eq9} 
P(V_{x, \;y, \;z} )=\frac{1}{\sqrt{\pi } \; V_{0} } e^{-V_{x, \;y, \;z\;}^{2} /V_{0}^{2} } , 
\end{equation} 

\noindent with the velocity spread $V_{0} =\sqrt{2k_{B} T/M} $ (here $k_{B} $ is the Boltzmann constant and \textit{M} is the atom mass). In order to suppress the Doppler effect on the power shifts we should satisfy $\delta _{1}^{*} \gg k_{1} V_{0} $ and $\delta _{3}^{*} \gg k_{3} V_{0} $; that is, the laser detunings from one-photon resonances should be much larger than the Doppler widths of these resonances. In these conditions the Doppler- and recoil-free three-photon excitation spectrum of Eq.~\eqref{Eq8} could be observed even for hot atom samples in vapor cells. In order to suppress the power shift it is also reasonable to choose $\Omega _{1} =\Omega _{3} $, $\delta _{2}^{*} =0$, and $\delta _{1}^{*} =-\delta _{3}^{*} $.

Third, the resulting excitation spectrum is obtained by averaging Eq.~\eqref{Eq8} over velocity distributions in particular atom samples. The value of $\rho_{33} $ averaged over velocity distribution represents a fraction of Rydberg atoms excited in an atom sample. The method of averaging depends on the interaction time of atoms with laser radiation at the crossing point of the three laser beams. In an ultracold Rb atom sample with $T=1\; \mu {\rm K}$ the most probable velocity is $V_{0} \approx 1.4\; {\rm cm/s}$. In the experiments on quantum information processing, an interaction time is limited by several microseconds in order to perform fast quantum gates [3], and the displacement of atoms is negligible compared to the typical diameters of focused laser beams ($10-100\; \mu {\rm m}$). In high-resolution spectroscopy the interaction time should be as long as the lifetimes of Rydberg states ($\tau _{R} =65-85\; \mu {\rm s}$ at the ambient temperature 300 K for the principal quantum number $n=50$ [17]). Even in this case the displacement is small for ultracold atoms. However, in a hot vapor-cell sample at room temperature $T=300\; {\rm K}$ the most probable velocity is $V_{0} \approx 240\; {\rm m/s}$, and the interaction time is limited by the time of flight through laser beams. For a 1 mm in diameter laser beam the interaction time is thus limited by $5-10\; \mu {\rm s}$ and it depends on the velocities of particular atoms. Therefore, Eq.\eqref{Eq8} should also be averaged over the interaction time that depends on atom velocity.

\section{Multiphoton excitation spectra}

In what follows we present the results of our numerical calculations using Eqs.~\eqref{Eq8} and \eqref{Eq9} for various temperatures of an atom sample and compare them with the calculations for the one- and two-photon Rydberg excitation schemes in Fig.~1(a). The analytical formulas for the excitation probability at these transitions were derived in the same manner as Eqs.~\eqref{Eq7} and ~\eqref{Eq8}. The counterpropagating radiations at 780 and 480 nm were taken for the two-photon transition.

Figure~2 shows the spectra of (a) one-photon, (b) two-photon, and (c) three-photon Rydberg excitations for various temperatures of an atom sample. The interaction time is 10 $\mu {\rm s}$ (Fourier-transform width 100 kHz) and the Rabi frequency for all transitions is $\Omega /(2\pi )=50$~kHz ($\pi $ laser pulse that should transfer all population to the Rydberg state). The detunings of the intermediate states at two- and three-photon transitions are 500 MHz (this value is taken to exceed the Doppler widths of the one-photon transitions at all temperatures used). The Rabi frequencies of the intermediate transitions are $\Omega _{1} /(2\pi )=\Omega _{2} /(2\pi )=7.07$ MHz for the two-photon transition, and $\Omega _{1} /(2\pi )=\Omega _{3} /(2\pi )=20$ MHz and $\Omega _{2} /(2\pi )=125$ MHz for the three-photon transition. These values require reasonable laser powers below 200 mW in the typical experimental conditions, according to our calculations of the dipole moments which have been confirmed to agree well with the experiment [18]. The lifetime of the Rydberg state is chosen to be $\tau _{R} =80\; \mu {\rm s}$. The detuning $\delta_2$ (1367~nm radiation) is varied for the three-photon transition, and detuning of the 480 nm radiation is varied for the two-photon transition.

All Rydberg excitation spectra in Fig.~2 demonstrate about 100 kHz Fourier-transform linewidth for the frozen atoms at $T=0\; {\rm K}$. The Rabi oscillations at the wings are due to the square shape of the laser pulse. Increasing the temperature to $T=1\; \mu {\rm K}$ results in partial decoherence of the one-photon excitation due to the Doppler broadening of 78 kHz. The two-photon transition is broadened by 19 kHz and is almost unaffected. At $T=10\; \mu {\rm K}$ the two-photon transition experiences decoherence from the Doppler broadening of 58 kHz, while the three-photon transition remains unaffected. The further increase of the temperature demonstrates substantial broadenings of the one-photon and two-photon transitions, in sharp contrast with the spectrum of the three-photon excitation. The latter has been found to be extremely robust against the heating of the atoms sample up to $T=10\; {\rm K}$. Even in a hot atom sample at $T=300\; {\rm K}$ the three-photon resonance is as narrow as the two-photon resonance at $T=50\; \mu {\rm K}$. 

\begin{figure}
\includegraphics[scale=0.43]{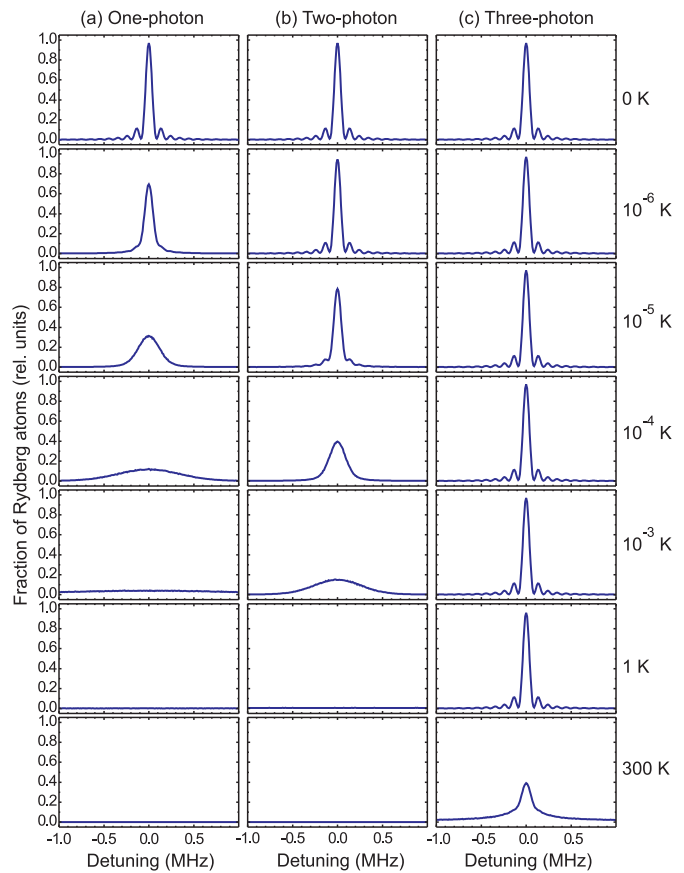}
\caption{\label{Fig2}(Color online) Spectra of (a) one-photon, (b) two-photon, and (c) three-photon laser excitation of Rydberg states in Rb atoms calculated for the interaction time $10\; \mu {\rm s}$ and Rabi frequency 50 kHz ($\pi $ laser pulse) for various temperatures of an atom sample. The detunings of the intermediate states at two- and three-photon transitions are 500 MHz. The lifetime of the Rydberg state is $80\; \mu {\rm s}$.}
\end{figure}

Figure~2 evidences a tremendous improvement of the spectral resolution at the three-photon Rydberg excitation compared to the one- and two-photon excitations. We may expect that using the geometry of Fig.~1(b) it is possible to substantially increase the accuracy of the spectroscopic measurements on three-photon transitions even in hot Rb vapor cells [14-16]. In particular, the hyperfine structure of lower Rydberg states might be resolved and measured. Although 1$-$2 MHz wide resonances of electromagnetically induced transparency have been demonstrated for the two-photon Rydberg excitation in Rb vapor cells [19] and cold atom samples [20], the three-photon method offers a potentially higher spectral resolution and signal-to-noise ratio, since almost all atoms in the atom sample are excited to Rydberg state, independently of their velocities.

\section{Coherence at multiphoton excitation}

The next issue to be considered is coherence of Rydberg excitation on a microsecond time scale, which is required for quantum information processing with Rydberg atoms [3]. Figure~3 presents the Rabi oscillations at (a) one-photon, (b) two-photon, and (c) three-photon laser excitation of Rb Rydberg states calculated at the line center and at the Rabi frequency $\Omega /(2\pi )=500$~kHz for various temperatures of an atom sample. The detunings of the intermediate states in the two- and three-photon transitions are 500 MHz. The Rabi frequencies of the intermediate transitions are $\Omega _{1} /(2\pi )=\Omega _{2} /(2\pi )=22.4$~MHz for the two-photon transition and $\Omega _{1} /(2\pi )=\Omega _{3} /(2\pi )=40$ MHz and $\Omega _{2} /(2\pi )=312.5$ MHz for the three-photon transition. The lifetime of the Rydberg state is $80\; \mu {\rm s}$.

\begin{figure}
\includegraphics[scale=0.43]{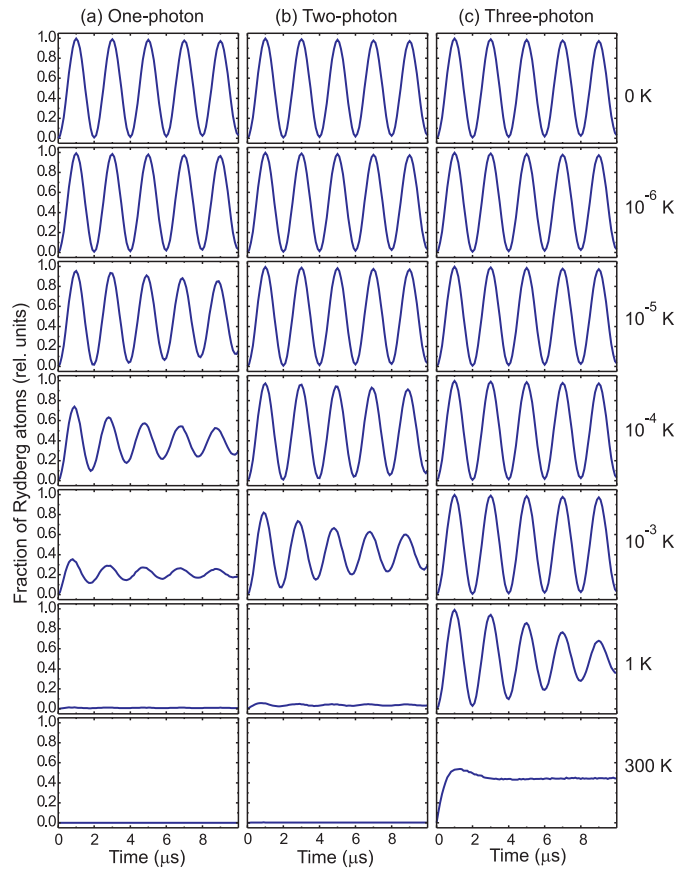}
\caption{\label{Fig3} (Color online) Rabi oscillations at (a) one-photon, (b) two-photon, and (c) three-photon laser excitation of Rydberg states in Rb atoms calculated at the line center and at the Rabi frequency 500 kHz for various temperatures of an atom sample. The detunings of the intermediate states in two- and three-photon transitions are 500 MHz. The lifetime of the Rydberg state is $80\; \mu {\rm s}$.}
\end{figure}

At $T=0\; {\rm K}$ a weak dephasing of the Rabi oscillations is observed in Fig.~3 at $t=10\; \mu {\rm s}$ due to the slow decay of the Rydberg state. Apart from that dephasing, the Rydberg excitation is coherent on a microsecond time scale up to $T=10\; \mu {\rm K}$ for the one-photon transition, up to $T=100\; \mu {\rm K}$ for the two-photon transition, and up to $T=1\; {\rm K}$ for our three-photon transition. Even at $T=300\; {\rm K}$ the three-photon Rydberg excitation exhibits the residuals of the Rabi oscillations, which are comparable to those observed experimentally at the two-photon transition in a cold Rb atom sample at $T\approx 100\; \mu {\rm K}$  [11]. We conclude that the three-photon Rydberg excitation in the geometry of Fig.~1(b) offers much longer coherence time and signal-to-noise ratio for the Rabi oscillations, again due to the fact that almost all atoms in the atom sample are excited to Rydberg state, independently of their velocities.

\begin{figure}
\includegraphics[scale=0.47]{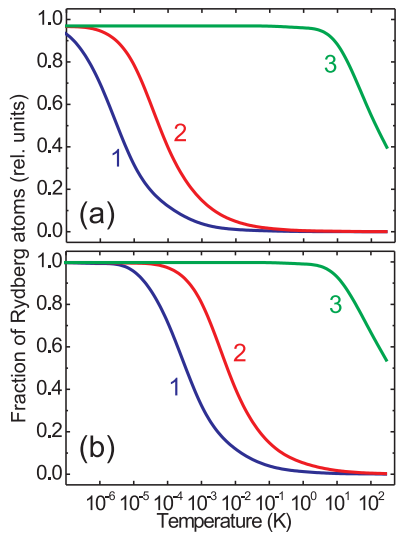}
\caption{\label{Fig4} (Color online) Probability of Rydberg excitation by $\pi$ laser pulse at the line center of the one-photon (curve 1), two-photon (curve 2), and three-photon (curve 3) transitions versus temperature of an atom sample for the Rabi frequency (a) 50 kHz (as in Fig.~2) and (b) 500 kHz (as in Fig.~3). The detunings and Rabi frequencies of the intermediate transitions are the same as in Figs.~2 and 3. The lifetime of the Rydberg state is $80\; \mu {\rm s}$.}
\end{figure}

In order to evaluate the effect of Doppler decoherence on Rydberg quantum gates, Fig.~4 presents the numerically calculated probability of Rydberg excitation by $\pi $ laser pulse at the line center of the one-photon, two-photon, and three-photon transitions versus temperature of an atom sample for the Rabi frequency (a) 50 kHz (as in Fig.~2) and (b) 500 kHz (as in Fig.~3). The detunings, Rabi frequencies of the intermediate transitions, and interaction times are the same as in Figs.~2 and 3. The lifetime of the Rydberg state is $80\; \mu {\rm s}$. It is seen that coherence on the three-photon transition weakly changes with the Rabi frequency compared to the one- and two-photon transition. The temperature dependence has a thresholdlike behavior with the threshold temperature corresponding to the equal Rabi frequency and Doppler width. At large Rabi frequency in Fig.~4(b), the error of Rydberg excitation by $\pi $ laser pulse on the three-photon transition at $T=100\; \mu {\rm K}$ is 0.3\%, completely due to the decay of the Rydberg state (otherwise, the error is zero to a precision better that $10^{-6} $). The error on the two-photon transition is 2.7\%. This indicates an order-of-magnitude improvement of the Rydberg gate error for the three-photon transition compared to the two-photon excitation.

\section{Discussion}

We have to make some comments on the shortcomings of the three-photon laser excitation. One drawback is that for the same Rabi frequency the three-photon excitation requires more laser power than one- and two-photon excitations. On the other hand, this power can be applied to the $5P_{3/2} \to 6S_{1/2} $ transition having a large dipole moment and therefore requiring only modest laser power below 100 mW. Another drawback is that the atom sample should be smaller than the diameters of the laser beams and tight focusing to increase the laser intensity cannot be applied for large atom samples. Finally, our particular three-photon excitation scheme allows us to excite only \textit{nP} Rydberg states. These states have Stark-tuned F\"orster resonances at $n\le 38$, which are suitable for quantum information processing  based on resonant dipole-dipole interaction of Rydberg atoms [21,22]. Another three-photon excitation scheme $5S_{1/2} \to 5P_{3/2} \to 5D_{5/2} \to nP,\; nF$ [14,15] offers also the excitation of the \textit{nF} states for which accidental F\"orster resonances can be expected at higher \textit{n} to increase the interaction energy. Other Rydberg states can also be excited in the weak dc electric field due to mixing of their wave functions and breakdown of the selection rules.

To conclude, the three-step laser excitation of Rydberg states by three different laser beams can be arranged in a starlike geometry that simultaneously eliminates the recoil effect at photon absorption and Doppler broadening. Our analytical and numerical calculations have shown that this approach works for both cold atom samples and hot vapors, if the intermediate states of the three-photon transition are detuned by more than respective single-photon Doppler widths. Compared to the one- and two-photon Rydberg excitation schemes, three-photon excitation offers much higher spectral resolution and longer coherence time and improves the signal-to-noise ratio due to the fact that almost all atoms in the atom sample are excited to Rydberg state, independently of their velocities. It can be especially useful for recoil-free Rydberg excitation in Bose-Einstein condensates, which otherwise are noticeably heated by one- or two-photon absorption [2]. Another application can be a study of the possible dynamical crystallization and supersolids in an ultracold blockaded Rydberg gas [23,24]. This method can also be suitable to the precision spectroscopy of low-excited atoms or molecules if appropriate three-photon transitions are available or laser-induced three-photon absorption and emission is used in \textit{N} schemes as in Refs.~[7,25,26].

\begin{acknowledgments}
This work was supported by the RFBR (Grants No.~10-02-00133 and No.~10-02-92624), by the Russian Academy of Sciences, by Presidential Grants No.~MK-6386.2010.2 and No.~MK-3727.2011.2, by the Dynasty Foundation, and by FP7-PEOPLE-2009-IRSES project "COLIMA".
\end{acknowledgments}

\end{document}